\documentclass[twocolumn]{aastex62}

\usepackage{savesym}
\savesymbol{tablenum}
\usepackage{siunitx}
\usepackage{amsmath}
\usepackage{graphicx}
\restoresymbol{SIX}{tablenum}
\shorttitle{The Jet Turnover Frequency}
\shortauthors{Hammerstein, G\"{u}ltekin, \& King}

\begin{document}

\title{Probing the Jet Turnover Frequency Dependence on Black Hole Mass and Mass Accretion Rate}

\author{Erica Hammerstein}
\affil{University of Michigan \\
1085 South University Avenue \\
Ann Arbor, MI 48109, USA}
\affil{University of Maryland \\
4296 Stadium Dr. \\
College Park, MD 20742, USA}

\author{Kayhan G\"{u}ltekin}
\affil{University of Michigan \\
1085 South University Avenue \\
Ann Arbor, MI 48109, USA}

\author{Ashley King}
\affiliation{ KIPAC, Stanford University \\ 
452 Lomita Mall \\ 
Stanford, CA 94305, USA}

\correspondingauthor{Erica Hammerstein}
\email{ekhammer@astro.umd.edu}

\begin{abstract}

We have examined a sample of 13 sub-Eddington supermassive black holes hosted by galaxies spanning a variety of morphological classifications to further understand the empirical fundamental plane of black hole activity. This plane describes black holes from stellar-mass to supermassive and relates the mass of an accreting black hole and its radio and X-ray luminosities. A key factor in studying the fundamental plane is the turnover frequency, the frequency at which the radio continuum emission  becomes optically thin. We measured this turnover frequency using new VLA observations combined, when necessary, with archival \textit{Chandra} observations. Radio observations are in the range of 5--40 GHz across four frequency bands in B-configuration, giving high spatial resolution to focus on the core emission. We use Markov Chain Monte Carlo methods to fit the continuum emission in order to find the turnover frequency. After testing for correlations, the turnover frequency does not display a significant dependence on either mass or mass accretion rate, indicating that more complicated physics than simple scaling and optical depth effects are at play, as has been suggested by recent theoretical work.
\end{abstract}


\section{Introduction}
\label{intro}

Scaling relations of black hole properties and activity have advanced our understanding of the similarities and differences between accreting systems across the black hole mass regime. Several scaling relations deal with a supermassive black hole (SMBH) and its host galaxy: $M_{\rm BH}$--$M_{ \rm bulge}$, $M_{\rm BH}$--$L_{\rm bulge}$, and $M_{\rm BH}$--$\sigma$. The first of these, $M_{\rm BH}$--$M_{ \rm bulge}$, measures the correlation between SMBH mass and host galaxy bulge mass \citep{kormendy95}. The second scaling relation is that of the black hole mass ($M_{\mathrm{BH}}$) to host galaxy bulge or spheroid luminosity \citep[ $L_{\mathrm{bulge}}$; e.g.,][]{dressler89,kormendy93,magorrian98}. The scatter in these two relations is much greater than in the third relation, $M_{\rm BH}-\sigma$ \citep{ferrarese00,gebhardt00a}, which relates the mass of the black hole and the velocity dispersion of the host galaxy's bulge or spheroid component and displays a remarkably tight correlation.

There are also other scaling relations, such as $\sigma^2_{\rm NXS}$--$M_{\rm BH}$ and $R_{\rm BLR}$--$L$, which deal with the immediate environment of the black hole. The first of these relations, $\sigma^2_{\rm NXS}$--$M_{\rm BH}$, relates the variability amplitude of the X-ray spectrum and the mass of the black hole, where the two are anti-correlated \citep[e.g.,][]{green93,lawrence93,nandra97}. The $R_{\rm BLR}$--$L$ correlation relates the radius of the broad line region to the quasar host galaxy continuum luminosity. This relation is useful in studying black hole masses and mass functions, the evolution and co-evolution of black holes and their host galaxies, and accretion in SMBHs \citep{bentz13}. 

In this paper, we focus on the relationship between black hole mass, X-ray luminosity, and radio luminosity that outline the ``fundamental plane of black hole activity'' (hereafter ``fundamental plane''). This relationship was first quantified in low/hard state black hole binaries by a tight non-linear correlation between radio and X-ray luminosity \citep{corbel03,gallo03}. These studies were followed up independently by both \citet{merloni03} and \citet{falcke04}, who further established a correlation between radio and X-ray luminosity in X-ray binaries (XRBs) as well as active galactic nuclei (AGN). \citet{plotkin12} explored the use of the fundamental plane as a tool in separating radiative processes around accreting black holes. They also further refined the coefficients of the fundamental plane and investigated the intrinsic scatter in the relation.

A key factor in studying the fundamental plane is the jet break frequency ($\nu_t$)\,---\,the frequency at which the slope of the non-thermal emission seen from the jet transitions from a flat power-law at frequencies lower than $\nu_t$ to a steep power-law at frequencies higher than $\nu_t$. Here we make the distinction between a flat power-law, with spectral index $\alpha \leq 0.40$, and a steep power-law, with spectral index $\alpha > 0.40$, where the flux density of the spectrum goes as $S_{\nu} \propto \nu^{-\alpha}$. This point is interpreted as a transition from optically thick synchrotron emission to optically thin synchrotron emission and has been proposed to scale with black hole mass, $M_{BH}$, by $\nu_t\propto {M_{BH}}^{-1}$ and with X-ray luminosity, $L_X$, at the turnover frequency by $\nu_t \propto L_X^{1/3}$ \citep{falcke95,markoff03}. We can measure $\nu_t$ by observing the source on either side of the expected break frequency and assume that the emission we see is synchrotron in both X-ray and radio bands.

Previous studies have attempted to understand the relation between $\nu_t$ and other black hole properties through studies of both XRBs and SMBHs. \citet{russell13} increased the number of observations of jet spectral breaks in XRBs. They also searched for a global relation between $\nu_{t}$ and luminosity but did not find one in the sample. However, they did find a correlation between the luminosity at the turnover frequency and the X-ray luminosity for hard state black hole X-ray binaries. More recently, \citet{russell14} examined how $\nu_t$ evolved over the course of an outburst in the black hole candidate X-ray binary MAXI J1836-194. Their results disagree with the previously proposed relationship $\nu_{\rm t} \propto L_{\rm X}^{1/3}$, with the luminosity of MAXI J1836−194 following a line almost perpendicular to the proposed relationship. However, they state that the proposed relation may hold true once the XRB has settled into a canonical hard state. By examining only one source, however, they were only able to test against changes in mass accretion rate, $\dot{m}$, of the system and necessarily cannot constrain the relationship between $M_{BH}$ and $\nu_t$. Other studies, such as \citet{ceccobello18}, have explored jet configurations using numerical modeling of accreting systems in hopes to better constrain the creation and structure of these jets and the origin of observed emission.

In order to better constrain scaling relationships between $\nu_t$, $\dot{m}$, and $M_{BH}$ we have examined a sample of thirteen accreting sub-Eddington SMBHs in a variety of galaxy classifications to further understand the fundamental plane of black hole activity and scaling relations between accreting stellar mass black hole systems and supermassive black hole systems. The configuration of these supermassive black holes is one where the accretion flow is in a radiatively inefficient state and there may be a characteristic jet-dominated accretion flow. This type of environment is similar to low-hard state X-ray binaries. Studying sources in such a configuration can yield important information on the disk--jet connection for accreting black holes of this and other Eddington rate configurations, most theories of which predict that jet production is highly dependent on mass accretion rate. In section \ref{sample} we present our sample of SMBHs in detail and motivate our experimental design. Section \ref{obs} details our observations, while section \ref{analysis} details our analysis and results. Finally, in section \ref{discussion} we discuss the implications that our results have on the proposed scaling relations for $\nu_t$ and the fundamental plane.

\section{Sample \& Design}
\label{sample}

We present our sample of thirteen SMBHs within host galaxies of a variety of galaxy classifications. Our sample is designed such that we may find any correlation between $\nu_{\rm t}$ and $\dot{m}$ or $M_{\rm BH}$ ($\nu_{\rm t} \propto \dot{m}^a \cdot {M_{\rm BH}}^b$, where $a$ and $b$ are some constants related to the proposed correlation). In order to test the turnover frequency dependence on both mass and mass accretion rate, we devised our sample to consist of two sub-samples so that we may test the dependence of the turnover frequency on only mass or mass accretion rate at one time. One sub-sample has eight sources with approximately constant mass accretion rate ($L_X/L_{\mathrm{Edd}} \sim 10^{-6}$) but varying black hole masses ($10^7$--$10^9 M_{\odot}$). A key ingredient in this study is the use of dynamically measured masses. Our sample of SMBHs has a mixture of stellar- and gas- dynamically measured masses \citep[with the exception of NGC 4258 which is a megamaser mass estimate;][]{herrnstein05}, eliminating systematic uncertainties created by secondary mass measurements.
The second sub-sample has eight sources with approximately constant black hole mass ($10^8 M_{\odot}$) but varying mass accretion rates ($L_X/L_{\mathrm{Edd}} \sim 10^{-6}$--$10^{-9}$). Since we cannot directly measure $\dot{m}$, we use $L_{\rm X}/L_{\rm Edd}$ as a proxy as there is a well-defined relationship between the bolometric correction and Eddington ratio \citep{vasudevan07}, which varies by less than an order of magnitude for Eddington ratios less than $\sim$0.1. Therefore, $L_{\rm X}/L_{\rm Edd}$ is a reasonable substitute for $\dot{m}$ as it is a direct observable and less complicated to obtain, although it is important to note that as stated by \citet{nagar05}, the fractional Eddington luminosity may underestimate the true accretion power. Three sources are in both sub-samples, having both $L_{\rm X}/L_{\rm Edd} \sim 10^{-6}$ and $M_{\rm BH}\sim 10^8 M_{\odot}$. Table \ref{tab:sampledata} lists these sources and indicates which sub-sample they are in along with several selected properties. 

One potential source of systematic uncertainty that remains with our sample is the use of SMBHs with masses measured using different techniques.  If, for example, gas-dynamical techniques tend to underestimate black hole masses beyond what is quantified in the calculated uncertainties, then this would compromise our analysis.  It is possible that gas-dynamical modeling that does not include non-Keplerian motion can introduce discrepancy between gas-dynamical and stellar dynamical measurements of the same system \citep{2018arXiv181005238J}.  Despite the potential for systematic discrepancies in individual cases \citep[e.g.,][]{gebhardt11, walsh13}, individual examination of four of the given gas dynamical measurements used in this work were deemed reliable by \citet{kormendy13} for scaling relation determination.  The use of mass measurements from both techniques (as well as with with megamaser measurements, which have been found to be consistent with stellar dynamical techniques [\citealt{siopis09}]), forms the basis of most studies of black hole mass scaling relations and secondary mass estimates \citep{2009ApJ...697..160B, 2013ApJ...764..184M, 2013ApJ...767..149B, 2018ApJ...864..146B}.  The one system not used for scaling relations by \citet{kormendy13} is NGC 4261.  As we discuss in sections \ref{results} and \ref{discussion}, our analysis of NGC 4261 only yields a turnover frequency lower limit, thus eliminating any potential concern for bias from systematic uncertainties in mass measurement.

\begin{deluxetable*}{lcccccccc}
\tablecolumns{8}
\tablecaption{AGN sample.}
\tablehead{\colhead{ID} & \colhead{Source} & \colhead{Distance} & \colhead{$\log({M_{\rm BH}/M_{\odot}})$} & \colhead{Galaxy}  & \colhead{$\log{(L_{\rm X}/L_{\rm Edd})}$} &  \colhead{Primary} & \colhead{Phase} & $\theta$ \\ & & \colhead{(Mpc)} & & Type & & Calibrator & Calibrator & \arcsec}

\startdata
1 & IC 1459\tablenotemark{b} & 28.92 & 9.39 $\pm$ 0.08 & E4 & $-$6.68 $\pm$ 0.12 & 3C48 & J2302$-$3718 & 2.66 \\
2 & NGC 1300\tablenotemark{b} & 21.50 & 7.88 $\pm$ 0.41 & SBbc & $-$6.19 $\pm$ 0.41 & 3C147 & J0312$-$1449 & 2.99 \\
3 & NGC 1316\tablenotemark{a} & 20.95 & 8.23 $\pm$ 0.07 & E4 & $-$7.31 $\pm$ 0.29 & 3C48 & J0334$-$4008 & 2.99\\
4 & NGC 3245\tablenotemark{a} & 21.38 & 8.38 $\pm$ 0.05 & S0 & $-$7.18 $\pm$ 0.22 & 3C286 & J1022+3041 & 2.86 \\
5 & NGC 3377\tablenotemark{a} & 10.99 & 8.25 $\pm$ 0.23 & E5 & $-$8.40 $\pm$ 0.30 & 3C286 & J1042+1203 & 0.86 \\
6 & NGC 3379\tablenotemark{a} & 10.70 & 8.62 $\pm$ 0.11 & E1 & $-$8.62 $\pm$ 0.20 & 3C286 & J1042+1203 & 0.86 \\
7 & NGC 3998\tablenotemark{ab} & 14.30 & 8.93 $\pm$ 0.04 & S0 & $-$5.62 $\pm$ 0.09 & 3C286 & J1146+5356 & 1.21\\
8 & NGC 4258\tablenotemark{b}& 7.27 & 7.58 $\pm$ 0.005 & SABbc & $-$4.83 $\pm$ 0.12 & 3C286 & J1219+4829 & 2.83\\
9 & NGC 4261\tablenotemark{ab} & 32.36 & 8.72 $\pm$ 0.09 & E2 & $-$5.84 $\pm$ 0.13 & 3C286 & J1222+0413 & 0.90\\
10 & NGC 4486\tablenotemark{b} & 16.68 & 9.79 $\pm$ 0.03 & E1 & $-$7.15 $\pm$ 0.10 & 3C286 & J1218+1105 & 0.94 \\
11 & NGC 4526\tablenotemark{a} & 16.44 & 8.65 $\pm$ 0.13 & S0 & $-$7.46 $\pm$ 0.16 & 3C286 & J1239+0730 & 0.88\\
12 & NGC 4594\tablenotemark{ab} & 9.87 & 8.82 $\pm$ 0.03 & Sa & $-$6.77 $\pm$ 0.21 & 3C286 & J1239$-$1023 & 1.19 \\
13 & NGC 5128\tablenotemark{b} & 3.63 & 7.76 $\pm$ 0.08 & E & $-$5.81 $\pm$ 0.14 & 3C147 & J1323$-$4452 & 0.26\\
\enddata
\tablecomments{The name of the source, the distance \citep{kormendy13}, mass of the central black hole \citep[and references therein]{gultekin09}, galaxy type \citep{kormendy13}, fractional Eddington luminosity \citep{gultekin09,gultekin12}, primary and phase calibrators for each source in our sample, and resolution $\theta$ obtained for each source are listed here in alphabetical order by source name. The first column, ID, is used in Figures \ref{fig:lxledd} and \ref{fig:mass} to identify the sources.}
\tablenotetext{\rm a}{Indicates that a source belongs to the sub-sample with constant black hole mass.} \tablenotetext{\rm b}{Indicates that a source belongs to the sub-sample with constant $L_{\rm X}/L_{\rm Edd}$.}
\label{tab:sampledata}
\end{deluxetable*}

\section{Observations with VLA}
\label{obs}

We observed the sample in a range of radio frequency bands, from 0.23 GHz to 40 GHz (P-, L-, C-, Ku-, and Ka-bands) using the NSF's Karl G. Jansky Very Large Array (VLA). Data were collected under the project code VLA/16A-245 in B-configuration from 2016 June 3 to 2016 September 24. Average scan time was approximately 270 seconds per band. We calibrated the data using the Common Astronomy Software Applications package (CASA; v5.0.0-218). First, we performed manual flagging to account for bad antennas, scans, and any other problematic observations. Next, we carried out antenna position corrections, initial flux density scaling, adopting the \citet{perley13} flux scale, and initial phase calibration. We then performed delay, bandpass, and gain calibrations. The primary calibrators, used for flux density and bandpass calibrations, as well as the phase calibrators for each source are listed in Table \ref{tab:sampledata}. We applied these calibrations to the observations of our sources and began initial imaging. In order to avoid resolving out flux at higher frequencies and creating an artificially steep radio spectrum, we  eliminated longer baselines at higher frequencies and shorter baselines at lower frequencies so that each observing band for a source has the same approximate resolution (the sources range in resolution between 0.26\arcsec -- 2.99\arcsec). 

We then split up each observing band into frequency bins. Frequency bins were created according to four spectral windows per bin, with four bins in the L-band and sixteen bins in the Ka-band. We imaged each frequency bin using the CLEAN algorithm within CASA, setting the gain to 0.1, and the threshold at which to stop cleaning to approximately 2.5 times the dirty map rms, which we determined from an obviously source-free region of the dirty map. We used the Clark PSF calculation mode with Briggs weighting and a robust parameter of 0.5. Unlike the other point sources, the emission from NGC 4486 consisted of core emission plus its prominent jet.  This, however, is not an issue in our analysis as we have high enough spatial resolution to focus only on the core emission in this source. We used only the emission consistent with the unresolved point source in obtaining flux densities. We discuss creation of spectra from this imaging in the next section.

\section{Analysis}
\label{analysis}
\subsection{Spectra}
\label{spectra}
We extracted the integrated flux density using CASA's imfit routine, which fits 2D Gaussian curves in the image plane, in order to construct a spectrum for each source. The errors on each spectral data point consist of the following added in quadrature: $\sigma_{\rm fit}$, the fit uncertainty calculated by imfit; $\sigma_{\rm flux}=3\%$, an absolute flux uncertainty \citep{perley13}; and $\sigma_{\rm rms}$, the cleaned map rms taken from an obviously source free region of each cleaned map. For each spectrum, we used emcee \citep{foreman13}, a Python implementation of the \citet{goodman10} affine-invariant Markov Chain Monte Carlo (MCMC) ensemble sampler, to fit broken power law functions to the radio data according to:
\begin{align*}
  S_{\nu} = \begin{cases}
      S_0 \left ( \frac{\nu}{10000~{\rm MHz}}\right)^{-\alpha_1} & \text{if $\nu \leq \nu_{\rm t}$} \\
      S_0 \left ( \frac{\nu}{10000~{\rm MHz}}\right)^{-\alpha_2} & \text{if $\nu > \nu_{\rm t}$}
    \end{cases},
\end{align*}
where $S_0$ is the 10 GHz value and $\alpha_1$ and $\alpha_2$ are the spectral indices before and after the turnover frequency, respectively. MCMC, in general, is a prime choice for fitting any data set as it provides a straightforward way of characterizing confidence intervals of parameters, can marginalize over possible nuisance parameters, and is useful in cases where error bars may be poorly estimated.
We used the 16th, 50th, and 84th percentiles of the samples in the marginalized distributions as our estimated values of the fit and errors. It is important to note that while this fitting routine is empirical, the functional form is based on the simple physical model described in section \ref{intro}. Additionally, while the transition from optically thick to optically thin may be a gradual one, we do not have the spectral resolution required to observe this gradual change and instead observe a broken power law.

In 3 of the data sets, it was apparent that the turnover frequency was outside the range of our observations. In these cases we combined the radio data with X-ray data from previous \emph{Chandra} observations of these sources to find a break in the spectrum. Four cases displayed a steep radio spectrum ($\alpha_{\rm R} > 0.40$) across the whole of our observations, implying that the turnover frequency occurs at frequencies lower than what we observed with VLA. These cases yield upper limits. Three sources required a broken power law fit to the radio data, implying that the turnover frequency occurs in the observed radio. Four sources exhibited flat radio spectra and X-ray data from which we were unable to make strong constraints on the continuum slope. Errors on the turnover frequency were calculated using standard propagation of error. See the Appendix for detailed tables of all sources.

\subsection{Results}
\label{results}

\begin{figure}
    \centering
    \includegraphics[width=\columnwidth]{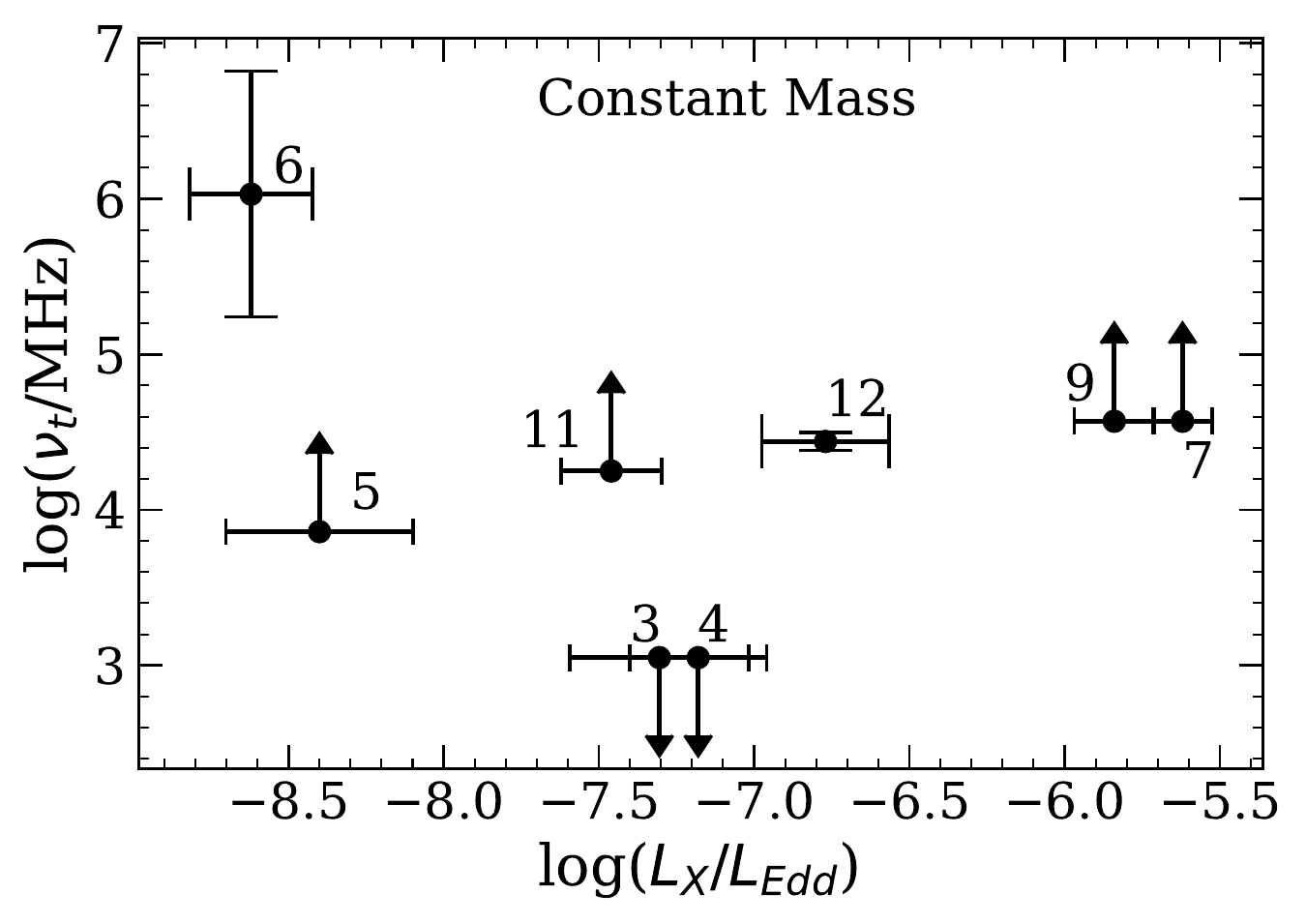}
    \caption{The turnover frequency as a function of fractional X-ray luminosity, which we use as a proxy for mass accretion rate. The sources in this figure all have approximately constant mass. Each source is given a unique number annotated on the figure that corresponds to the ID column in Table \ref{tab:sampledata}, consistent across Figures \ref{fig:lxledd} and \ref{fig:mass}. Correlation testing by assigning arbitrary extreme values for limits reveals a weak positive correlation. With a very high $p$-value, however, this result is not statistically significant.}
    \label{fig:lxledd}
\end{figure}

\begin{figure}
    \centering
    \includegraphics[width=\columnwidth]{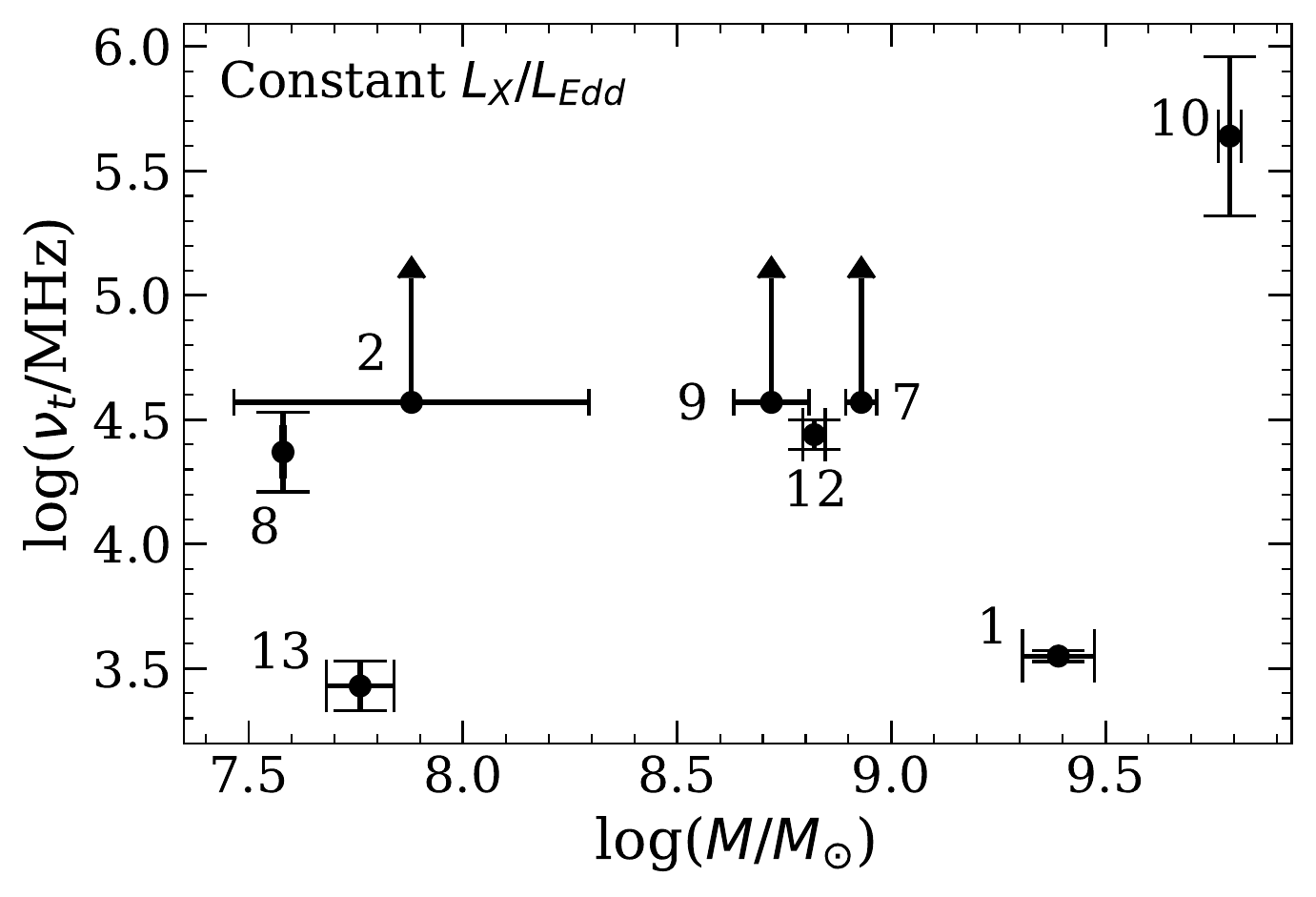}
    \caption{The turnover frequency as a function of mass of the central supermassive black hole. The sources in this figure all have approximately constant $L_{X}/L_{\rm Edd}$. Each source is given a unique number annotated on the figure that corresponds to the ID column in Table \ref{tab:sampledata}, consistent across Figures \ref{fig:lxledd} and \ref{fig:mass}. First inspection of this figure reveals a possible slight positive correlation and correlation testing confirms this. This result is not statistically significant, as testing also yields a very high $p$-value.}
    \label{fig:mass}
\end{figure}

We present the results of our observations on our sample of SMBHs and exploration of the turnover frequency dependence on several parameters for each source. These parameters include the mass of the central black hole and the mass accretion rate, for which we use the proxy $L_{X}/L_{\rm Edd}$. Figures \ref{fig:lxledd} and \ref{fig:mass} display the turnover frequency as a function of these values, neither of which displays an obvious trend. Table \ref{tab:results} lists values for $\alpha$ and $\nu_{t}$ for each source as well.

First, we explore potential trends in $\nu_{t}$ as a function of $L_{X}/L_{\rm Edd}$. Figure \ref{fig:lxledd} shows the turnover frequency as a function of $L_{X}/L_{\rm Edd}$, where $L_{X}/L_{\rm Edd}$ ranges from approximately $10^{-8.5}$--$10^{-5.5}$ and $\nu_{t}$ ranges from approximately $10^3$--$10^6$ MHz. First inspection of Figure \ref{fig:lxledd} may reveal a weak negative correlation, however, with only two detections and six upper or lower limits, a relation is hard to determine.

We tested for correlation using the Spearman rank-order correlation coefficient. Using a Monte Carlo technique, we were able to include upper and lower limits by testing many random combinations of these data points. This is justified as under our model assumptions, the turnover frequency must appear somewhere in the spectrum, and the precise value of each limit does not matter because we use rank-order correlation tests. Correlation testing reveals a weak positive correlation, with $r_S$ of 0.10. This is, however, not statistically significant with a $p$-value of 0.74. 

Next, we explore potential trends in $\nu_{t}$ as a function of $M_{\rm BH}$. Figure \ref{fig:mass} shows the turnover frequency plotted as a function of mass of the central SMBH, where $M_{\rm BH}$ ranges from approximately $10^{7.5}$--$10^{10} M_{\scriptscriptstyle\odot}$ and $\nu_t$ ranges from approximately $10^3$--$10^6$ MHz. This sub-sample has five detections and three upper or lower limits. Inspection of Figure \ref{fig:mass} reveals a possible slight positive correlation. 

First inspection of Figure \ref{fig:mass} reveals a possible positive correlation and testing  does reveal a weak positive correlation between the two properties with $r_S = 0.12$. Again, this is not statistically significant with a $p$-value of 0.74. Implications of these results are discussed in the next section.

\begin{deluxetable*}{lccc}
\tablecolumns{4}
\tablecaption{Results of fits.}
\tablehead{\colhead{Source} & \colhead{$\alpha_1$} & \colhead{$\alpha_2$} & \colhead{$\log{(\nu_{t})}$}}
\startdata
IC 1459	& $-$0.08 $\pm$ 0.08 & 0.73 $\pm$ 0.01 & 3.55 $\pm$ 0.02 \\
NGC 1300 & 0.27 $\pm$ 0.15 & \dots & $\geq$ 4.57 \\
NGC 1316 & \dots & 0.86 $\pm$ 0.02 & $\leq$ 3.05 \\
NGC 3245 & \dots & 0.50 $\pm$ 0.02 & $\leq$ 3.05 \\
NGC 3377 & $-$1.05 $\pm$ 0.96 & \dots & $\geq$ 3.86 \\
NGC 3379 & $-$0.10$\pm$ 0.07 & 1.05 $\pm$ 0.43 & 6.03 $\pm$ 0.79 \\
NGC 3998 & 0.33 $\pm$ 0.02 & \dots & $\geq$ 4.57 \\
NGC 4258 & 0.02 $\pm$ 0.04 & 0.45 $\pm$ 0.16 & 4.37 $\pm$ 0.16 \\
NGC 4261 & 0.04 $\pm$ 0.02 & \dots & $\geq$ 4.57 \\
NGC 4486 & 0.19 $\pm$ 0.01 & 1.22 $\pm$ 0.2 & 5.64 $\pm$ 0.32 \\
NGC 4526 & $-$0.32 $\pm$ 0.04 & \dots & $\geq$ 4.25 \\
NGC 4594 & $-$0.04 $\pm$ 0.04 & 5.19 $\pm$ 0.63 & 4.44 $\pm$ 0.06 \\
NGC 5128 & $-$0.55 $\pm$ 0.11 & 0.63 $\pm$ 0.12 & 3.44 $\pm$ 0.1
\enddata
\tablecomments{The columns are source name, $\alpha_1$, $\alpha_2$, and $\log{(\nu_{t})}$. Here $\alpha_1$ is the spectral index at frequencies lower than $\nu_{t}$ and $\alpha_2$ is the spectral index at frequencies higher than $\nu_{t}$. Sources are listed in alphabetical order.}
\label{tab:results}
\end{deluxetable*}

\section{Discussion}
\label{discussion}
In Section \ref{results} we found no significant correlation between $\nu_{t}$ and $M_{\rm BH}$ or $L_{X}/L_{\rm Edd}$ at low Eddington fraction. One might expect a correlation between these black hole properties. If jet power, $Q_j$, depends on accretion flow properties, then any connection between X-ray and radio properties should produce some sort of correlation. There is certainly no shortage of accretion inflow-outflow models, including radiatively inefficient models such as: advection-dominated accretion flows \citep[ADAF;][]{narayan94}, or radiatively inefficient accretion flows \citep[RIAF;][]{yuan03}, the adiabatic inflow-outflow solution \citep[ADIOS;][]{blandford99,begelman12}, and the convection-dominated accretion flow \citep[CDAF;][]{narayan00,quataert00,abramowicz02,igumenshchev02}. Near-Eddington accreting black holes are dominated by thermal emission \citep{shakura73}. \citet{markoff03} suggest that black holes accreting at sub-Eddington rates make a transition to a radiatively inefficient state and becomes dominated by non-thermal synchrotron emission from a jet. This gives rise to a predicted relationship between accretion power and the spectrum of a black hole's jets. \citet{merloni03} provide one model of a relationship between X-ray and radio properties that requires a radiatively inefficient accretion flow paired with a scale invariant synchrotron emitting jet. They do not, however, make any predictions for relationships between the turnover frequency and black hole properties. \citet{falcke04} expand on the \citet{markoff03} model for an X-ray/radio correlation and include consideration of optical depth effects, making $\nu_{\rm t}$ dependent on $L_{X}/L_{\rm Edd}$ and $M_{\rm BH}$, which is the primary investigation in this paper.

The origin of the radio emission in these galaxies must also be examined. \citet{ulvestad01} conclude that 
the source of the characteristic flat power law radio spectrum is consistent with synchrotron self-absorption. They also suggest that this flat power law must consist of several self-absorbed components, which determine the value of the spectral index. While this may be the case, our study searches for the last regime in which the spectrum transitions from optically thick to optically thin, as the spectrum will not be optically thick for frequencies out to infinity.

Our test of the proposed scaling relations does have several small limitations. One model of the jet is that all emission, including X-ray emission, comes from scattering of synchrotron emission by energetic electrons \citep{markoff03}. Thus, $L_{X}/L_{\rm Edd}$ may not directly probe the mass accretion rate. However, if jet power is correlated with the mass accretion rate, then $L_{X}/L_{\rm Edd}$ will still correlate well with mass accretion rate. Further study on the zone for electron acceleration in these jets is needed to explore this model. Additionally, if spin contributes to the observational properties of these systems, our small sample may be affected if there is a wide variety of spin parameters. As stated by \citet{sikora07}, fueling of AGN in spiral galaxies is likely not linked to galaxy mergers, and may be limited by several smaller accretion events fueled by molecular clouds with randomly oriented angular momentum vectors. It is probable that large ellipticals have experienced at least one major merger in their lifetimes, which are followed by accretion events that spin-up the black holes to consistently high values in these types of galaxies. Within our sample all but three sources are classified as early type galaxies meaning most of our sources likely have very similar spin parameters. However, it is also possible that all spirals have persistent accretion and thus spin-up their black holes and have similar high spin distributions while ellipticals may have very chaotic accretion that will randomly spin-up or spin-down their black holes. The environment of a particular AGN may also play a role in how collimated or suppressed a jet is, where a more dense environment (perhaps in a group or cluster) will confine a jet more than one found in the field. Despite these modest limitations, our study is the first to investigate $\nu_{t}-M_{\rm BH}$ using dynamical mass measurements.

As we have found no correlation between $\nu_{t}$ and $M_{\rm BH}$ or $L_{X}/L_{\rm Edd}$ for our sample of SMBHs, it is clear that more complicated physics is involved in these types of systems. This is further illustrated by more recent observational and theoretical work. \citet{russell14} showed that the model presented by \citet{falcke04} is not well followed in XRBs either, through their analysis of the outburst of MAXI J18360-94, which is able to follow $\dot{m}$ over a single black hole mass. Of course, detailed study of an individual XRB cannot probe mass dependent relationships, which our study is able to do. Our sample of SMBHs does not follow the predicted scaling of $\nu_{t}$ with $M_{\rm BH}$. This suggests that a fundamental piece of physics is missing in our current model and that optical depth effects in the jets may not be as straightforward as previously thought. It may be that jet geometry is different than our current understanding as well. Numerical studies, such as \citet{ceccobello18}, can investigate the role that optical depth effects play in the structure and observation of jets.

\section{Summary \& Conclusions}
A more tightly constrained relationship could help to predict masses of black holes if radio and X-ray properties are known. This would be extremely helpful in discovering and classifying intermediate mass black holes, which is a more challenging area of black hole masses to study. Future work will require further examination of the expected relationships across both the mass and mass accretion rate scale, including both stellar mass black holes and SMBHs in many different accretion states.

We have observed a sample of sub-Eddington accreting supermassive black holes in both radio and X-ray wavelengths in order to determine each source jet turnover frequency. We used MCMC techniques to fit the spectra of these observations and find this spectral break frequency, the turnover frequency. We then explored the turnover frequency dependence on mass and mass accretion rate to come to the following conclusions:
\begin{itemize}
    \item We have shown that the turnover frequency, $\nu_{t}$, does not display any significant correlation with black hole mass.
    \item We have also shown that the turnover frequency does not display any significant correlation with mass accretion rate.
    \item The lack of correlation may indicate that the simple jet structure we have assumed is not correct. Instead of a smoothly distributed collection of zones, the jet may be more clumpy than previously thought, or perhaps more or less compact. Both of these factors will have an effect on the turnover frequency.
\end{itemize}

\acknowledgments
The National Radio Astronomy Observatory is a facility of the National Science Foundation operated under cooperative agreement by Associated Universities, Inc. The authors thank Kristina Nyland and Sera Markoff for their helpful comments on improving this paper. \software{CASA (v5.0.0-218; \citep{mcmullin07}), emcee \citep{foreman13}}

\bibliographystyle{aasjournal}
\bibliography{turnover}

\appendix
In this Appendix we show a spectrum of each source, as indicated by the title of the legend in each figure.

\begin{figure}[ht]
    \centering
    \includegraphics[width=0.4\columnwidth]{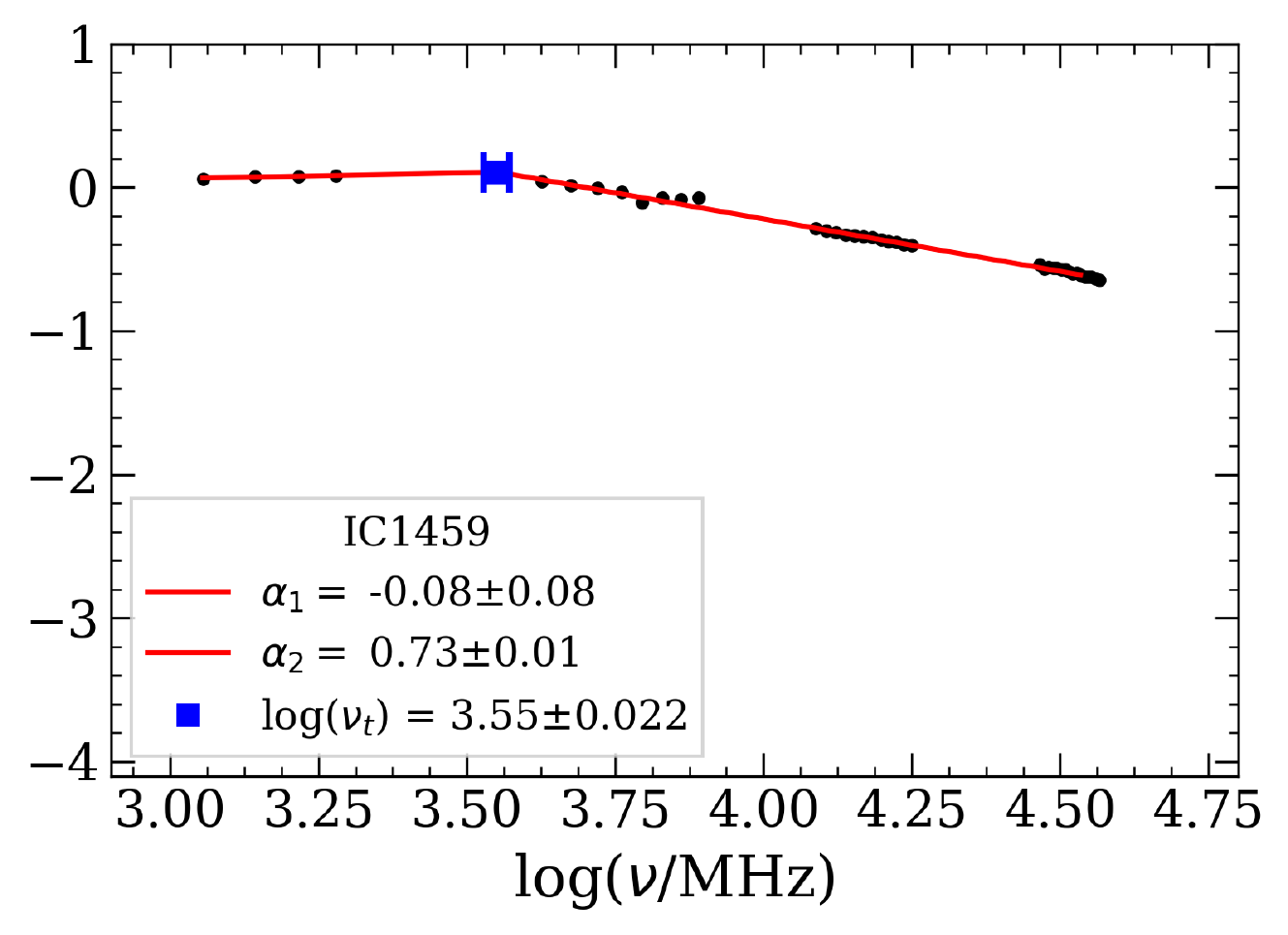}
    \includegraphics[width=0.4\columnwidth]{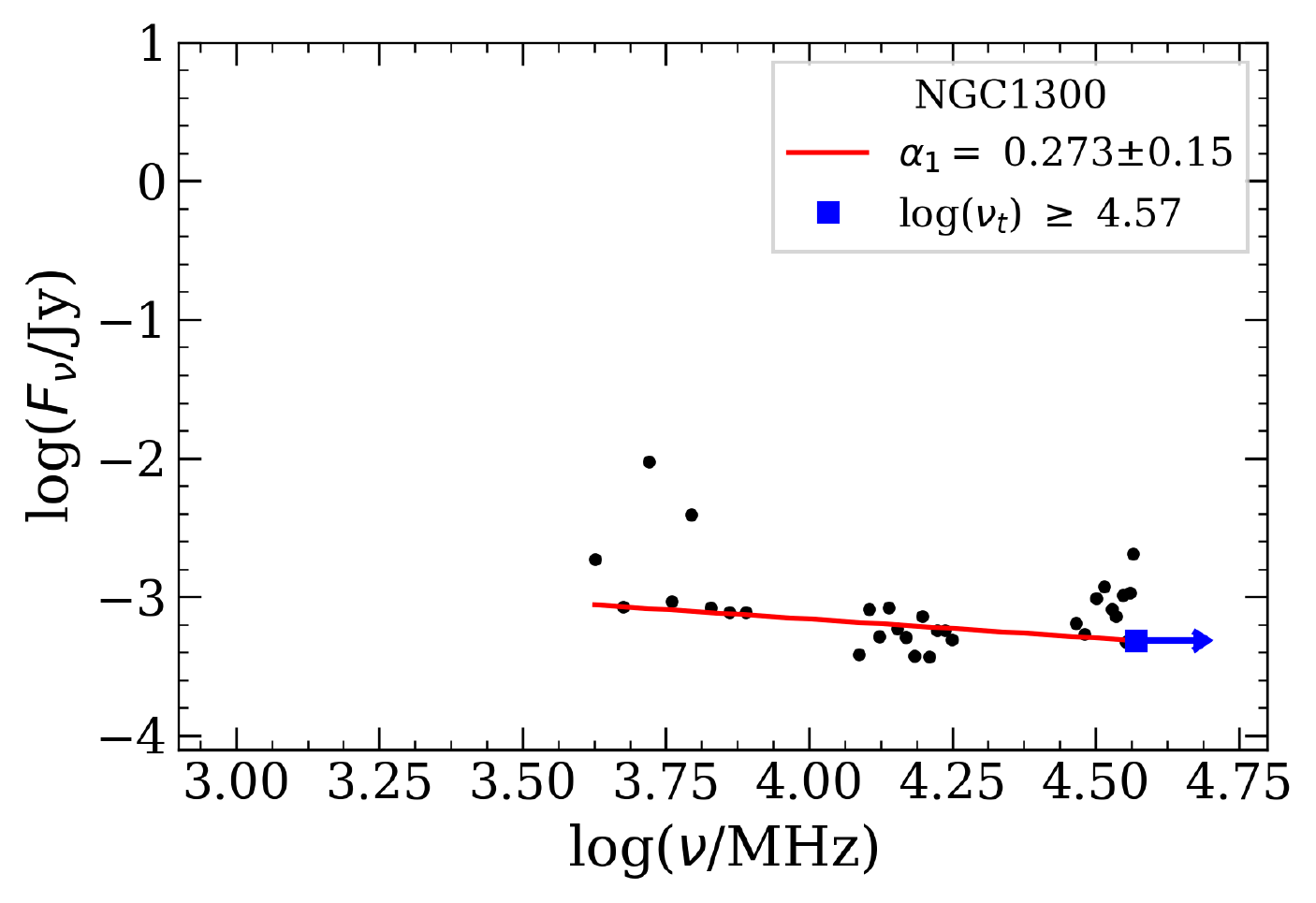}\\
    \includegraphics[width=0.4\columnwidth]{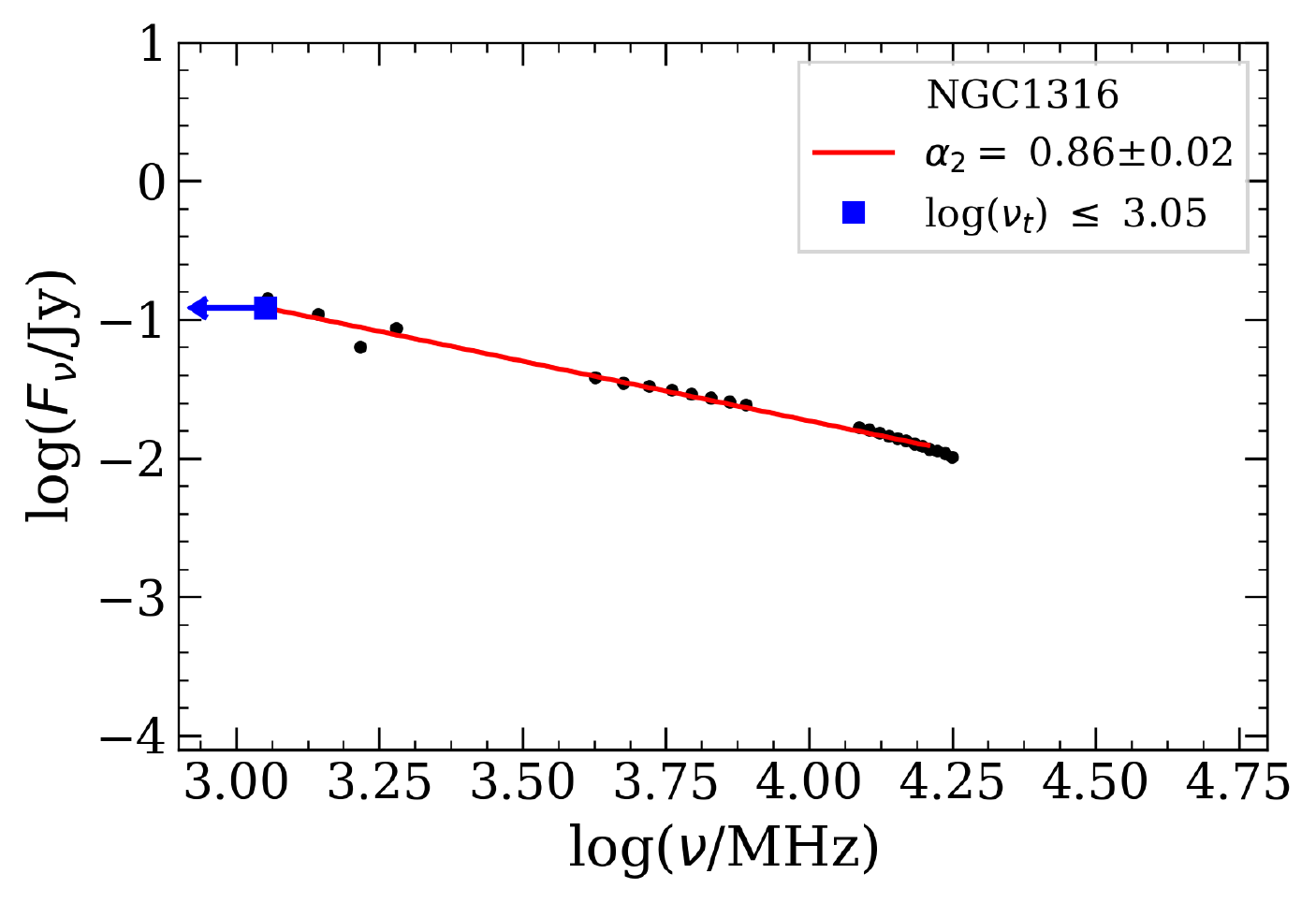}
    \includegraphics[width=0.4\columnwidth]{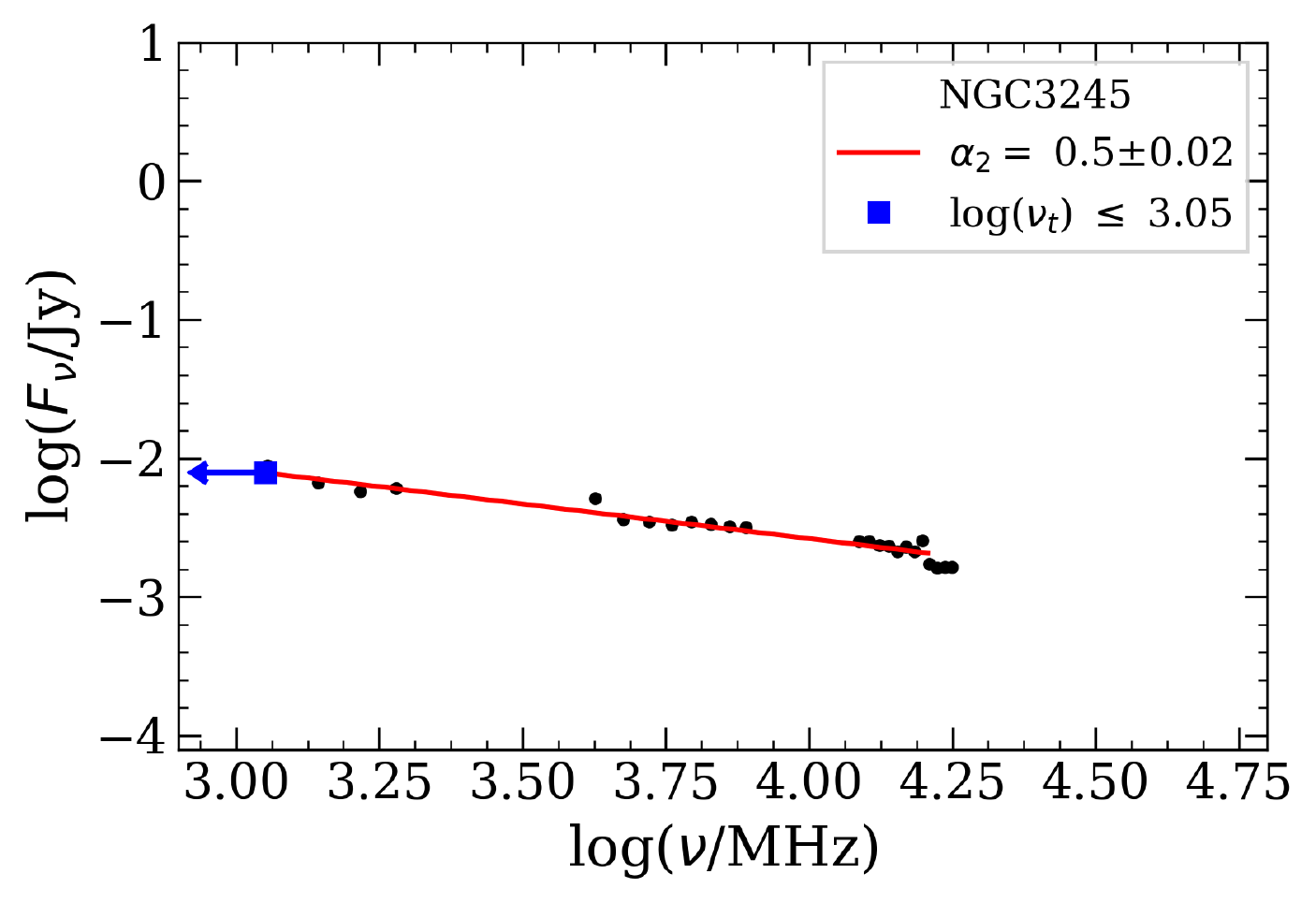}\\
    \includegraphics[width=0.4\columnwidth]{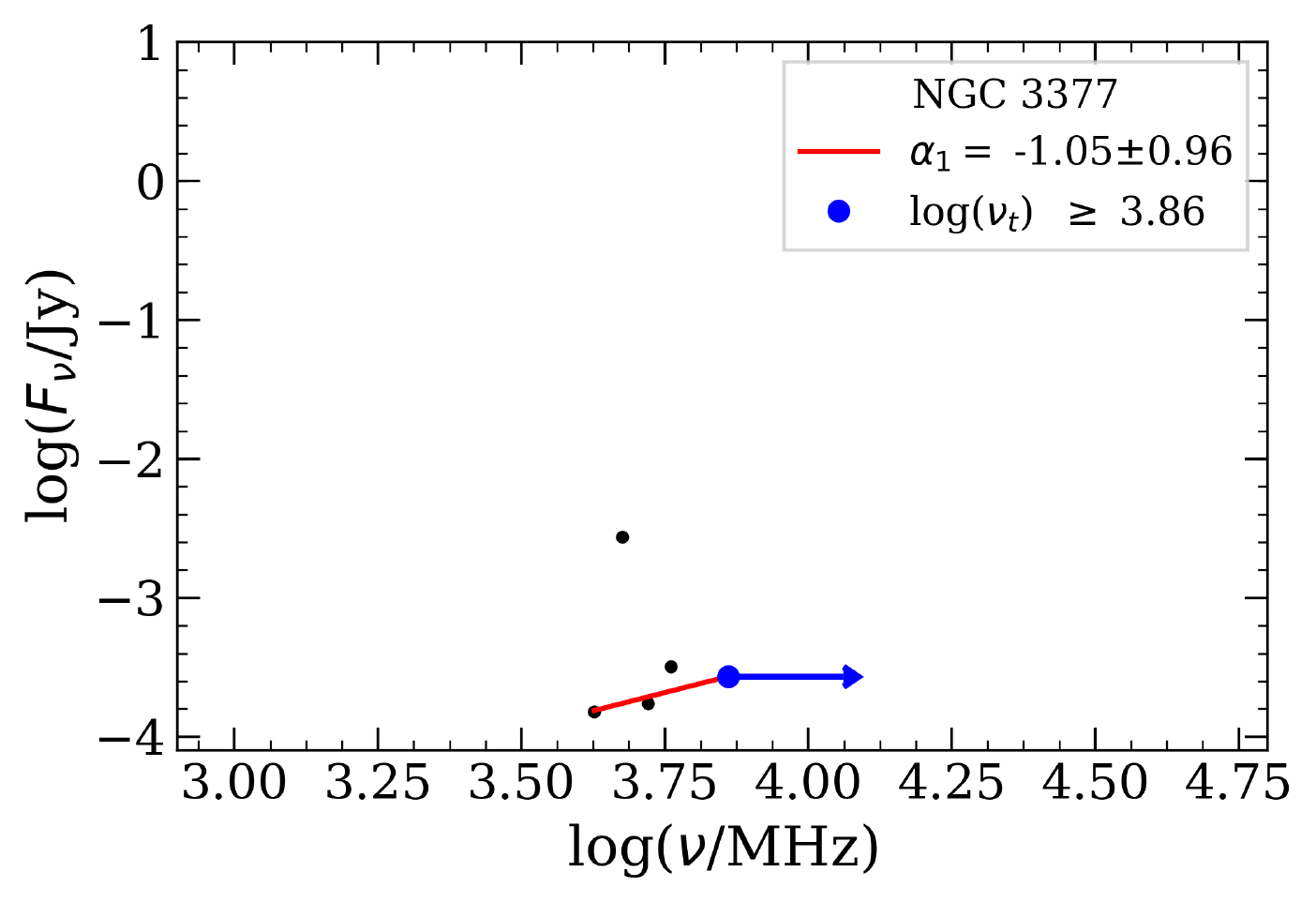}
    \includegraphics[width=0.4\columnwidth]{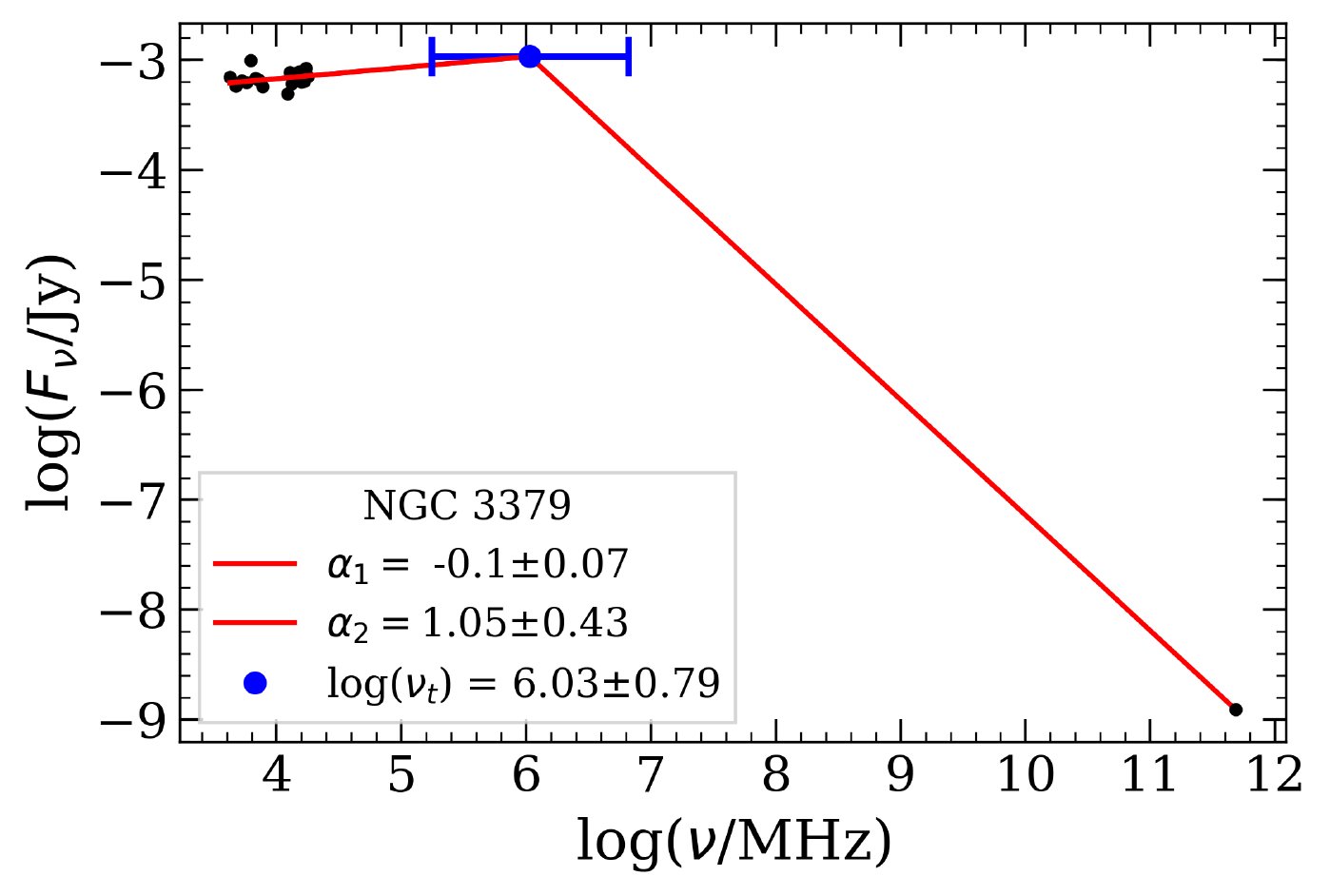}
    \caption{This figure shows spectra of IC1459, NGC1300, NGC1316, NGC3245, NGC3377, and NGC3379, as shown in the legend of each subpanel. The red lines indicate fits to the spectral data points, the values of which are noted in the legend as $\alpha_1$ or $\alpha_2$, where $\alpha_1$ and $\alpha_2$ are the spectral indices before and after the turnover frequency, respectively. The turnover frequency is indicated by a blue box, with error bars or upper and lower limits where appropriate. Subpanels that do not include X-ray data use the same scale for the axes.}
    \label{fig:spectra1}
\end{figure}

\begin{figure}[ht]
    \centering
    \includegraphics[width=0.4\columnwidth]{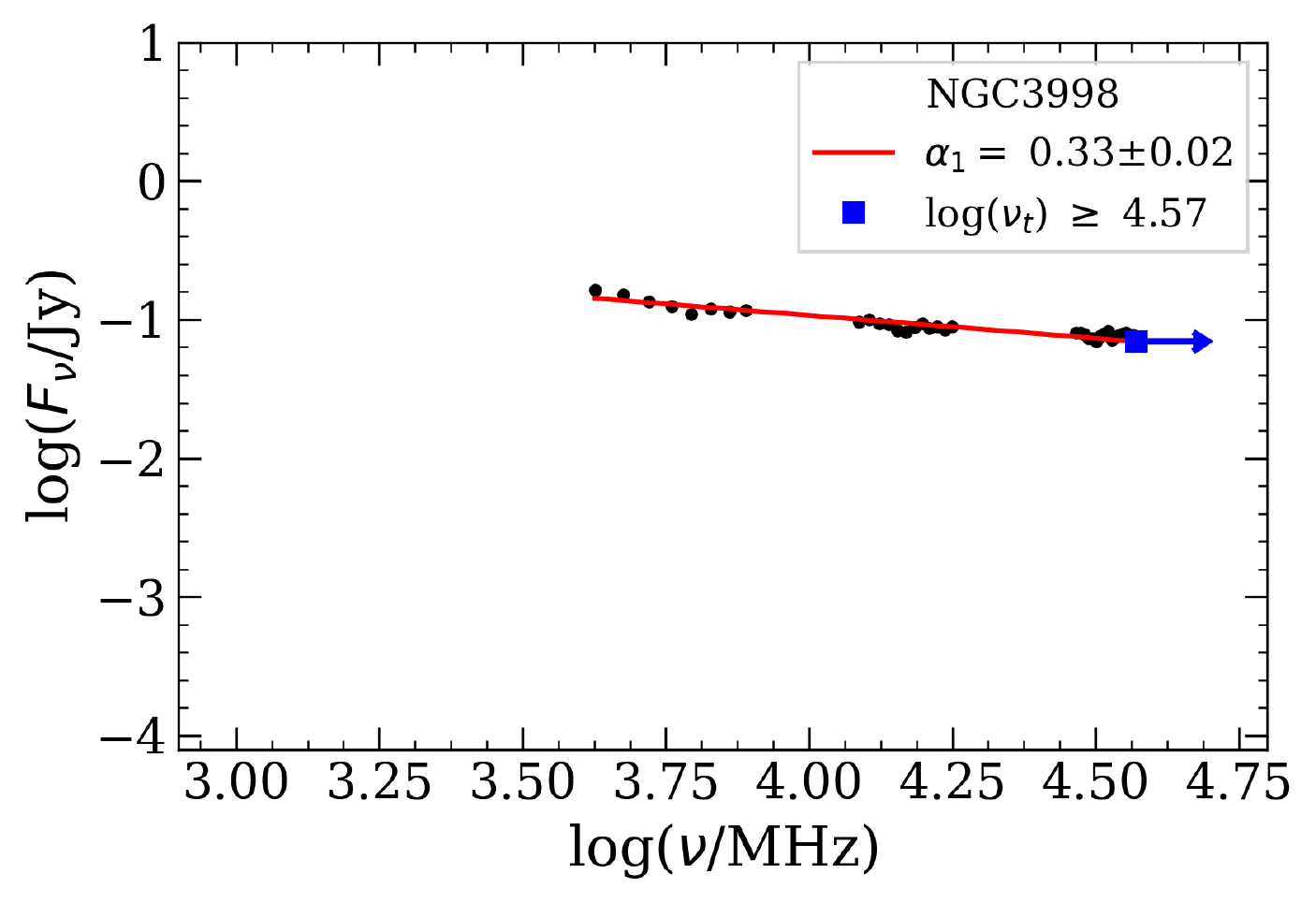}
    \includegraphics[width=0.4\columnwidth]{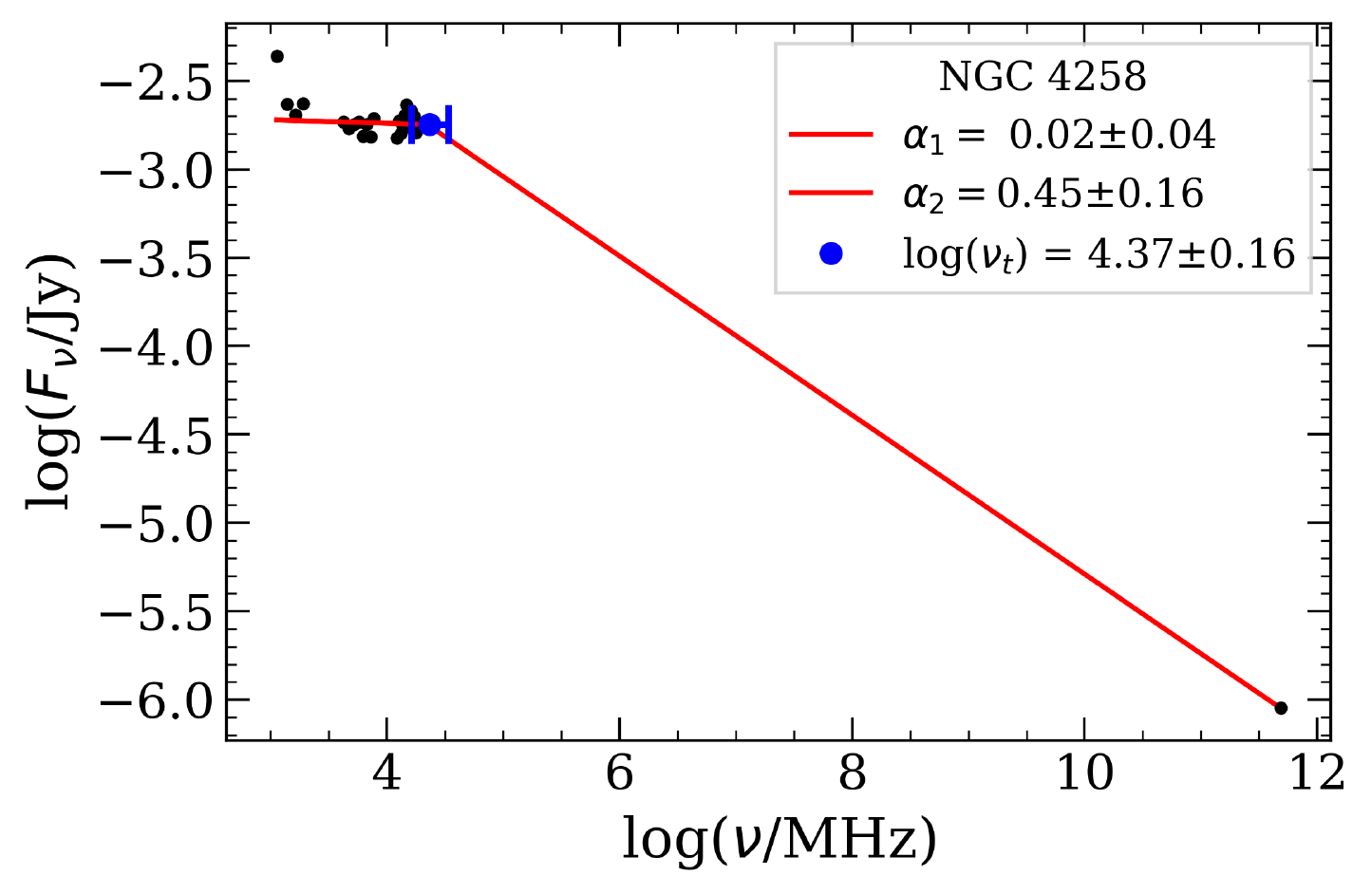}\\
    \includegraphics[width=0.4\columnwidth]{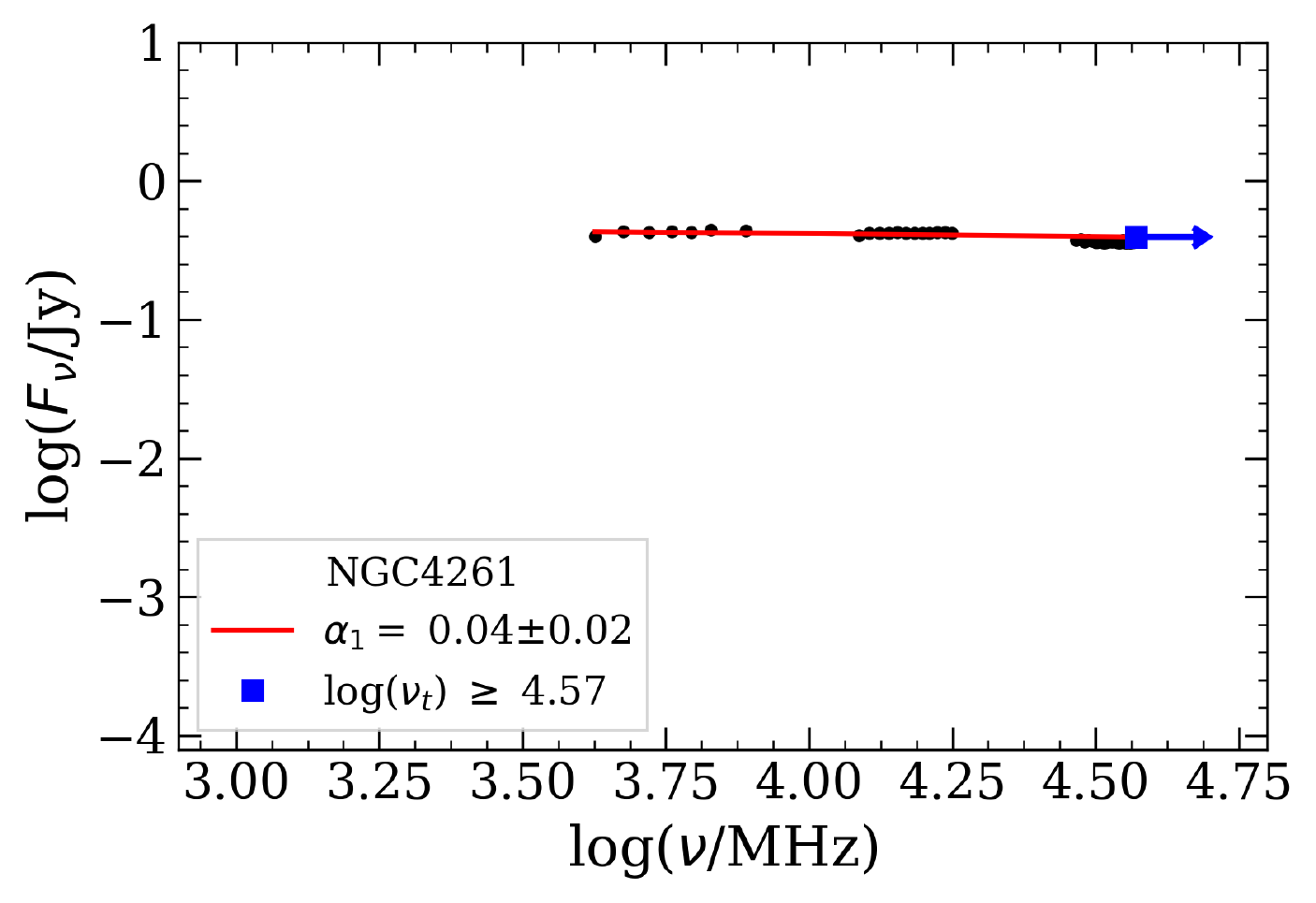}
    \includegraphics[width=0.4\columnwidth]{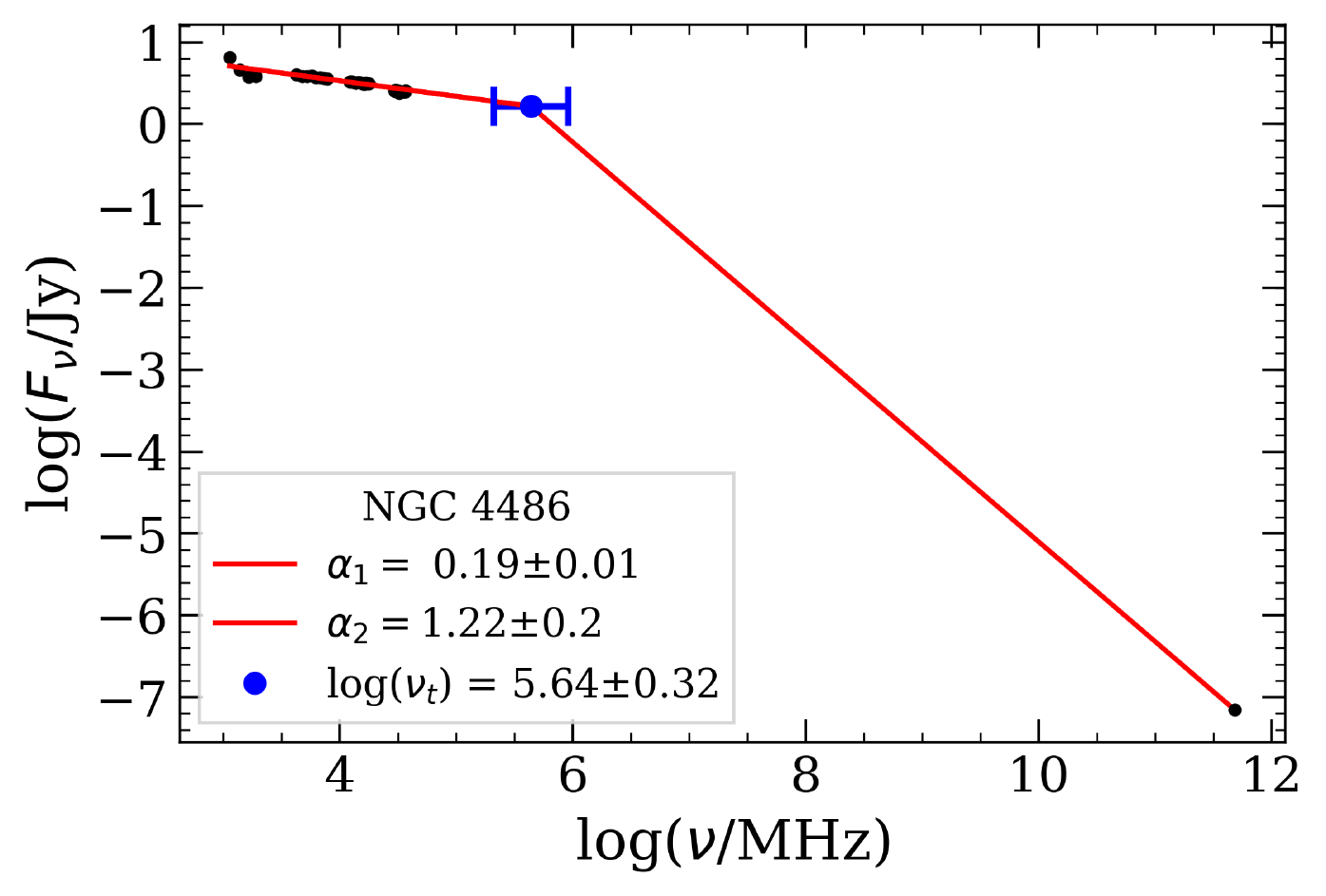}\\
    \includegraphics[width=0.4\columnwidth]{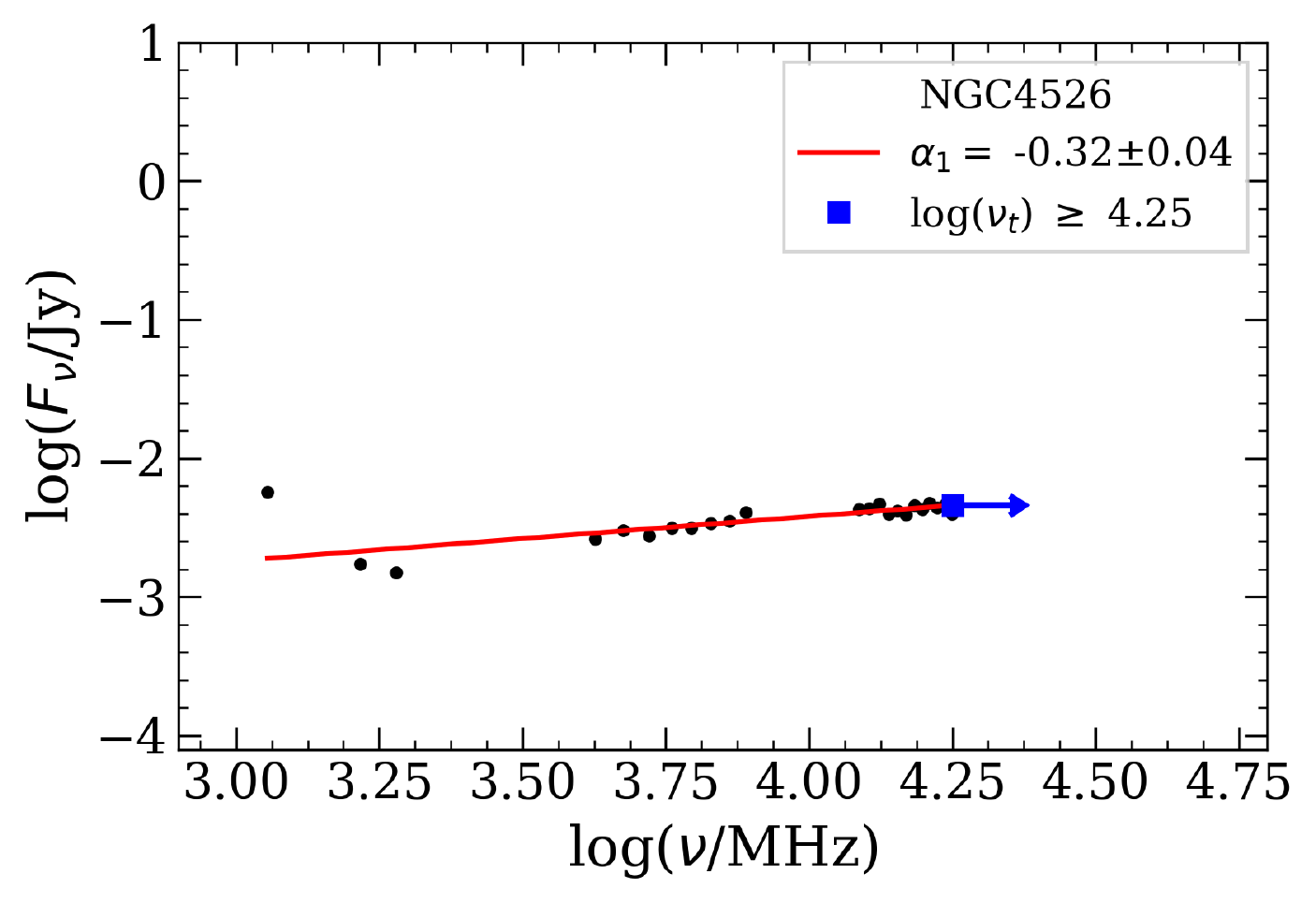}
    \includegraphics[width=0.4\columnwidth]{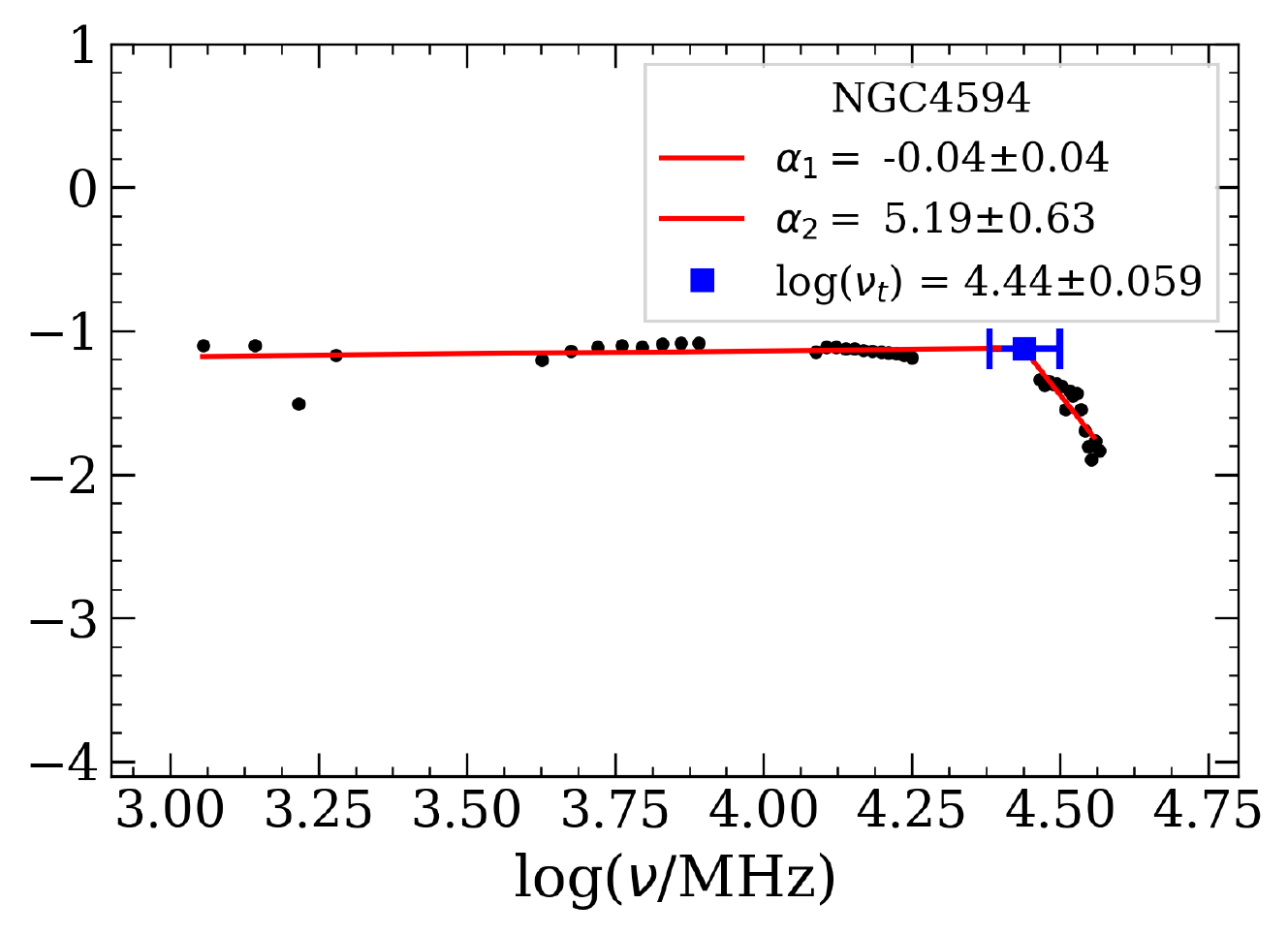}\\
    \includegraphics[width=0.4\columnwidth]{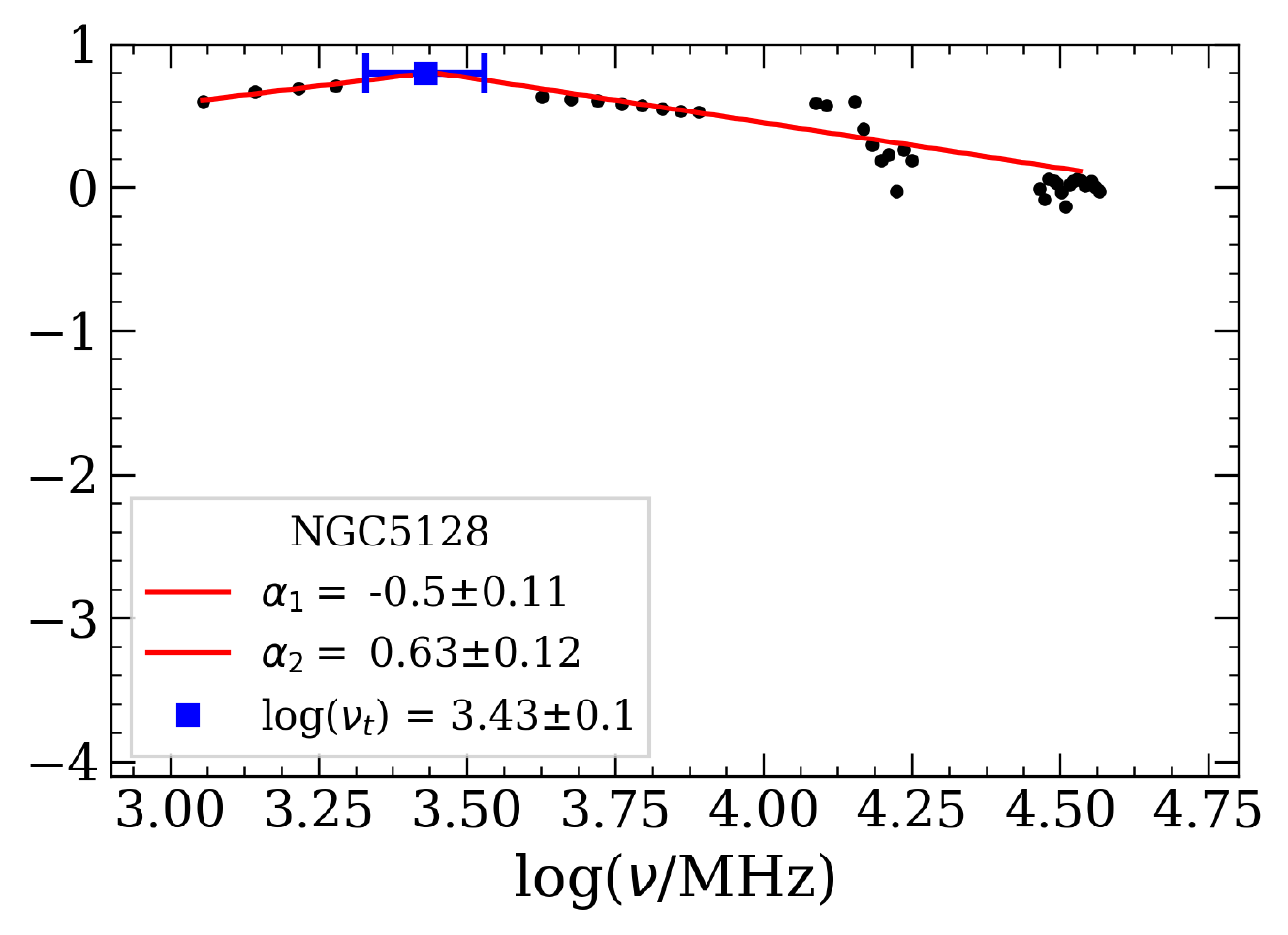}
    \caption{Same as Fig.\ \ref{fig:spectra1} but for NGC3998, NGC4258, NGC4261, NGC4486, NGC4526, NGC4594, and NGC5128.}
    \label{fig:my_label_continued}
\end{figure}

\end{document}